\documentclass[aps,prl,reprint,superscriptaddress,floatfix]{revtex4-1}

\usepackage{physics}
\usepackage{amsmath,amssymb,upgreek}

\usepackage[pdftex]{graphicx}
\DeclareGraphicsExtensions{.pdf,.jpeg,.png,.jpg}
\usepackage{hyperref}

\usepackage{soul}

\usepackage[utf8]{inputenc}

\begin{document}

\title{Spatial Noise Correlations in a Si/SiGe Two-Qubit Device from Bell State Coherences}
\date{\today}

\author{Jelmer M. Boter}
\thanks{These authors contributed equally to this work.}

\author{Xiao Xue}
\thanks{These authors contributed equally to this work.}

\author{Tobias S. Kr\"ahenmann}

\author{Thomas F. Watson}
\affiliation{QuTech and Kavli Institute of Nanoscience, Delft University of Technology, Lorentzweg 1, 2628 CJ Delft, The Netherlands}

\author{Vickram N. Premakumar}

\author{Daniel R. Ward}

\author{Donald E. Savage}

\author{Max G. Lagally}

\author{Mark Friesen}

\author{Susan~N.~Coppersmith}
\thanks{Present address: School of Physics, University of New South Wales, Sydney NSW 2052, Australia.}

\author{Mark A. Eriksson}

\author{Robert Joynt}
\affiliation{University of Wisconsin-Madison, Madison, WI 53706, USA}

\author{Lieven M.~K. Vandersypen}
\thanks{To whom correspondence should be addressed: \href{mailto://l.m.k.vandersypen@tudelft.nl}{l.m.k.vandersypen@tudelft.nl}}
\affiliation{QuTech and Kavli Institute of Nanoscience, Delft University of Technology, Lorentzweg 1, 2628 CJ Delft, The Netherlands}
\affiliation{Components Research, Intel Corporation, 2501 NE Century Blvd, Hillsboro, OR 97124, USA}

\begin{abstract}
We study spatial noise correlations in a Si/SiGe two-qubit device with integrated micromagnets.
Our method relies on the concept of decoherence-free subspaces, whereby we measure the coherence time for two different Bell states, designed to be sensitive only to either correlated or anti-correlated noise respectively. 
From these measurements, we find weak correlations in low-frequency noise acting on the two qubits, while no correlations could be detected in high-frequency noise.
A theoretical model and numerical simulations give further insight into the additive effect of multiple independent (anti-)correlated noise sources with an asymmetric effect on the two qubits.
Such a scenario is plausible given the data and our understanding of the physics of this system.
This work is highly relevant for the design of optimized quantum error correction codes for spin qubits in quantum dot arrays, as well as for optimizing the design of future quantum dot arrays.
\end{abstract}

\maketitle


Large-scale quantum computers will need to rely on quantum error correction (QEC) to deal with the inevitable qubit errors caused by interaction with the environment and by imperfect control signals. 
The noise amplitude can vary from qubit to qubit and furthermore can exhibit correlations or anti-correlations between qubits.
Most QEC error thresholds, such as the 1\%-threshold for the surface code~\cite{Wang2011}, are derived under the assumption of negligible correlations in qubit errors.
Other approaches such as decoherence-free subspaces (DFSs)~\cite{Lidar1998} are designed under the assumption of correlated noise, taking advantage of symmetry considerations to reduce the qubit sensitivity to external noise.
Examples for quantum dot based qubits include the singlet-triplet qubit~\cite{Levy2002,Petta2005} and the quadrupole qubit~\cite{Friesen2017}.
In addition, QEC schemes exist that can deal with short-range correlations in the noise \cite{Preskill2013}.
Spatial noise correlations have therefore been studied extensively, both theoretically~\cite{Rivas2015,Szankowski2016,Paz-Silva2017,Postler2018,Kwiatkowski2018,Premakumar2018,Krzywda2019} and experimentally~\cite{Monz2011,Postler2018,Ozaeta2019}.

Semiconductor quantum dots are promising hosts for spin qubits in quantum computation \cite{Loss1998}, because of their favorable scaling and excellent coherence properties.
Silicon, in particular, has excellent properties for long-lived spin qubits: intrinsic spin-orbit coupling is weak and hyperfine interaction is small~\cite{Zwanenburg2013}.
The hyperfine interaction can even be reduced further by isotopic purification.
In addition, silicon quantum dot fabrication is largely compatible with conventional CMOS industry, which allows large-scale manufacturing of silicon spin qubits and on-chip integration of classical control electronics~\cite{Vandersypen2017}.
In recent years, significant progress has been made with silicon spin qubits, showing tens of milliseconds coherence times~\cite{Veldhorst2014}, high-fidelity single-~\cite{Veldhorst2014,Kawakami2016,Yoneda2018} and two-qubit gates~\cite{Xue2019,Huang2019}, quantum algorithms~\cite{Watson2018}, strong spin-photon coupling~\cite{Samkharadze2018,Mi2018} and long-distance spin-spin coupling~\cite{Borjans2019}.

The most important decoherence sources in natural silicon quantum dots are the hyperfine interaction with nuclear spins and charge noise.
Nuclear spin noise is typically uncorrelated between adjacent dots~\cite{Chekhovich2013}.
Charge noise is usually caused by distant fluctuating charges~\cite{Jung2004,Paladino2014,Beaudoin2015}, which is expected to lead to spatial correlations on the length scale of interdot distances of 100 nm or less.
In the presence of a magnetic field gradient, which is commonly used for qubit selectivity and fast qubit control, qubits are sensitive to electric field fluctuations and charge noise will impact spin coherence~\cite{Kha2015,Kawakami2016}.
However, a quantitative measurement of spatial noise correlations in an actual two-qubit device is lacking.

Here we study experimentally spatial noise correlations in a Si/SiGe two-qubit device, by preparing Bell states in either the parallel or the anti-parallel subspace, similarly to recent work with NV centers in diamond~\cite{Bradley2019}. Via a Ramsey-style experiment, we find that Bell states in the anti-parallel subspace show a $\sim$30\% longer dephasing time than those in the parallel subspace. A Hahn-echo style measurement reveals no detectable difference in the decay time for the respective Bell states. We present a simple model to describe noise correlations on two qubits, including asymmetric noise amplitudes acting on the two qubits, and study numerically the combined effect of multiple (anti-)correlated, asymmetric noise sources. We use these simulations to assess which combinations of noise sources are compatible with the observed coherence times.


\begin{figure}[!htb]
	\centering
	\includegraphics{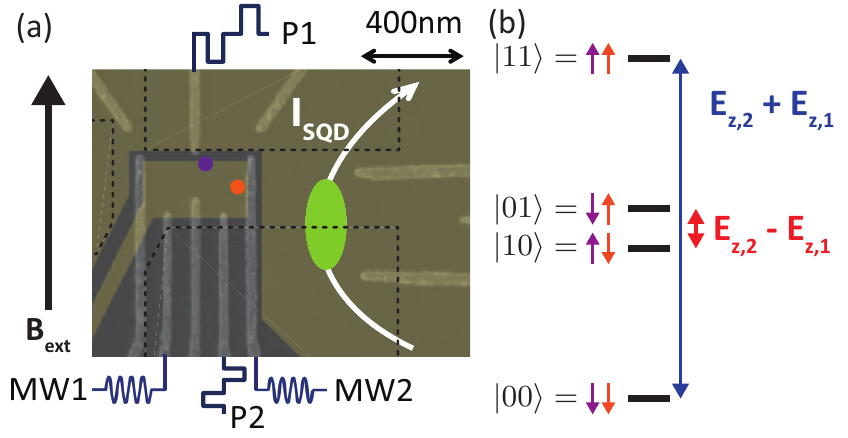}
	\caption{(a) Scanning electron micrograph of a similar Si/SiGe device as used in the measurements, showing the depletion gates used to define the potential landscape in the 2D electron gas accumulated by the yellow shaded gates (drawn digitally). 
	Purple and orange circles indicate the estimated positions of the two dots, occupied by one electron each, and the ellipse indicates a sensing quantum dot. 
	Two-qubit operations are controlled via gate voltage pulses applied to gates P1 and P2, and microwave signals for single-qubit control are applied to gates MW1 and MW2. The contours of cobalt micromagnets are indicated by the dashed black lines. 
	(b) Energy level diagram for two qubits in an inhomogeneous magnetic field, giving rise to a difference in Zeeman energy between the two qubits.}
	\label{fig:fig1}
\end{figure}

Figure~\ref{fig:fig1}(a) shows a schematic of the device used in this work, which is the same as described earlier~\cite{Watson2018,Xue2019}.
It comprises an electrostatically defined double quantum dot (DQD) in a two-dimensional electron gas (2DEG).
The 2DEG is confined in a 12-nm-thick silicon quantum well, 37 nm below the surface of an undoped Si/SiGe heterostructure with natural isotope composition.
On top of the heterostructure, we fabricate two gate layers with cobalt micromagnets.
The device is cooled down to $T\approx30$ mK and subject to an external magnetic field of B\textsubscript{ext} = 617 mT.
Suitable voltages are applied to accumulation and fine gates (in the top and bottom layer, respectively) to form a DQD in the single-electron regime.
Single-electron spin states are Zeeman split by the total magnetic field, and used to encode two single-spin qubits.
The micromagnets ensure individual qubit addressability by a gradient in the longitudinal magnetic field, resulting in spin resonance frequencies of 18.35 GHz and 19.61 GHz for qubit 1 (Q1) and qubit 2 (Q2), respectively. 

Figure~\ref{fig:fig1}(b) shows the resulting energy level diagram for the two qubits.
For perfectly correlated noise, fluctuations in the Zeeman energy for both qubits are the same: $\delta E_{Z,1} = \delta E_{Z,2} = \delta E_{Z}$.
Consequently, the sum of the two qubit energies fluctuates, $\Delta(E_{Z,1} + E_{Z,2}) = 2\delta E_Z$, while their difference is not affected, $\Delta(E_{Z,1} - E_{Z,2}) = 0$.
On the other hand, for perfectly anti-correlated noise $\delta E_{Z,1} = -\delta E_{Z,2}$, and the opposite holds for the sum and difference energies.
Therefore, an anti-parallel Bell state, which evolves in time at a rate proportional to the difference of the single-qubit energies, will be affected by anti-correlated noise, but not by correlated noise. A parallel Bell state, which evolves in time at a rate proportional to the sum of the single-qubit energies, is sensitive to correlated noise, but not to anti-correlated noise.
Such properties are exploited in DFSs and are used here as a probe for spatial correlations in the noise acting on the qubits.

Real systems are often subject to both uncorrelated and (anti-)correlated noise.
Furthermore, the noise amplitudes acting on different qubits are generally different, regardless of whether the noise is uncorrelated or (anti-)\\correlated.
We wish to capture all these scenarios in one unified theoretical formalism.
We include pure dephasing only, which is justified by the long $T_1$ times for spin qubits compared to the experiment and coherence timescales, and assume a quasi-static Gaussian joint probability distribution for the noise acting on the two qubits.
We can then express the two-qubit coherence times for an anti-parallel ($\ket{\Psi}  = (\ket{\downarrow\uparrow} -i \ket{\uparrow\downarrow})/\sqrt{2}$) and a parallel ($\ket{\Phi} = (\ket{\downarrow\downarrow} -i \ket{\uparrow\uparrow})/\sqrt{2}$) Bell state quantitatively as follows (see Supplemental Material~\cite{SuppMat}):
\begin{equation}
\begin{split}
    \left(\frac{1}{T_{2,\ket{\Psi}}^*}\right)^2 &= 2\pi^2 \left(\sigma_1^2 + \sigma_2^2 - 2\rho\sigma_1\sigma_2 \right), \\
    \left(\frac{1}{T_{2,\ket{\Phi}}^*}\right)^2 &= 2\pi^2 \left(\sigma_1^2 + \sigma_2^2 + 2\rho\sigma_1\sigma_2 \right),
\end{split}
\label{eq:T2s}
\end{equation}
where $\sigma_i^2$ is the variance of the noise in the resonance frequency of qubit $i$ (the single-qubit coherence time is given by $\left(\frac{1}{T_{2,i}^*}\right)^2 = 2\pi^2 \sigma_i^2$), and $\rho$ is a correlation factor ($-1 \leq \rho \leq 1$).
Positive $\rho$ indicates correlations, while negative $\rho$ indicates anti-correlations. 

\begin{figure}[!htb]
	\centering
	\includegraphics{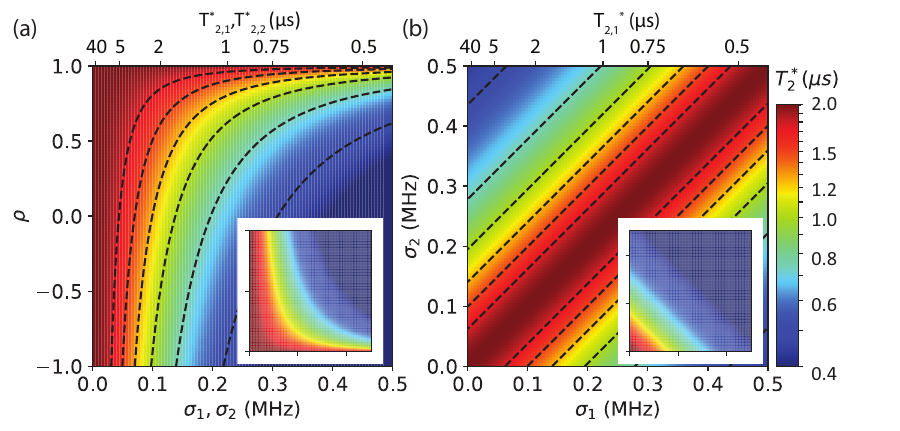}
	\caption{$T_{2,\ket{\Psi}}^*$ extracted from Eq.~\ref{eq:T2s} (a) as a function of correlation factor $\rho$ and noise amplitude $\sigma_1=\sigma_2$, and (b) as a function of $\sigma_1$ and $\sigma_2$ for $\rho=1$. Insets show the corresponding images for $T_{2,\ket{\Phi}}^*$. Contours correspond to (0.5, 0.75, 1.0, 1.25, 1.5, 1.75) $\mu$s. In all images an uncorrelated noise contribution corresponding to a Bell state coherence time of $2.0 \,\mu$s is added to prevent singularities.}
	\label{fig:fig2}
\end{figure}

The effect of the noise amplitudes $\sigma_i$ and the correlation factor $\rho$ on the coherence time for the anti-parallel Bell state $T_{2,\ket{\Psi}}^*$ is visualized in Fig.~\ref{fig:fig2}(a).
Here $\sigma_1=\sigma_2$, so for $\rho=1$, $\ket{\Psi}$ forms a true DFS and the noise has no effect regardless of its amplitude. 
With decreasing $\rho$, $T_{2,\ket{\Psi}}^*$ decreases, as the noise becomes initially less correlated ($\rho > 0$), then uncorrelated ($\rho=0$) and eventually anti-correlated ($\rho < 0$).
For $\rho=-1$, $T_{2,\ket{\Psi}}^*$ is only one fourth of the single-qubit coherence times.
For $T_{2,\ket{\Phi}}^*$ the corresponding image is mirrored around $\rho=0$, see the inset of Fig.~\ref{fig:fig2}a, and the longest coherence time occurs for $\rho=-1$.
Figure~\ref{fig:fig2}(b) shows the effect of asymmetric noise amplitudes on the two qubits for $\rho=1$. We see that despite the maximal correlation factor, a true DFS only exists for symmetric noise ($\sigma_1=\sigma_2$) and $\ket{\Psi}$ decoheres when $\sigma_1\neq\sigma_2$. 
Clearly, both the asymmetry in the noise and the correlation factor impact the two-qubit coherence.

From Eq.~\ref{eq:T2s}, we see that, as anticipated, experimental measurement of the decay times for the parallel and anti-parallel Bell states reveals whether (anti-)correlations in the noise acting on the two qubits are present. In order to quantify the correlation factor $\rho$, measurements of the single-qubit decay time are needed as well.  
We now summarize the experimental procedure; for more information on the measurement setup and individual qubit characteristics, see the Supplemental Material~\cite{SuppMat} and Ref.~\cite{Watson2018}.
Q2 is initialized and read out via spin-selective tunneling to a reservoir~\cite{Elzerman2004}.
Initialization of Q1 to its ground state is done by fast spin relaxation at a hotspot~\cite{Srinivasa2013}, and read-out of Q1 is performed by mapping its spin state onto Q2 via a controlled-rotation (CROT) gate followed by spin read-out of Q2 \cite{Watson2018}.
For single-qubit driving we exploit an artificial spin-orbit coupling, induced by cobalt micromagnets, for electric dipole spin resonance (EDSR)~\cite{Pioro-Ladriere2007}.
The two-qubit gate relies on the exchange interaction between the two qubits, controlled by gate voltage pulses. We operate in the regime where the Zeeman energy difference between the two qubits exceeds the two-qubit exchange interaction strength, hence the native two-qubit gate is the controlled-phase gate~\cite{Meunier2011,Veldhorst2015,Watson2018}.

\begin{figure}[!htb]
	\centering
	\includegraphics{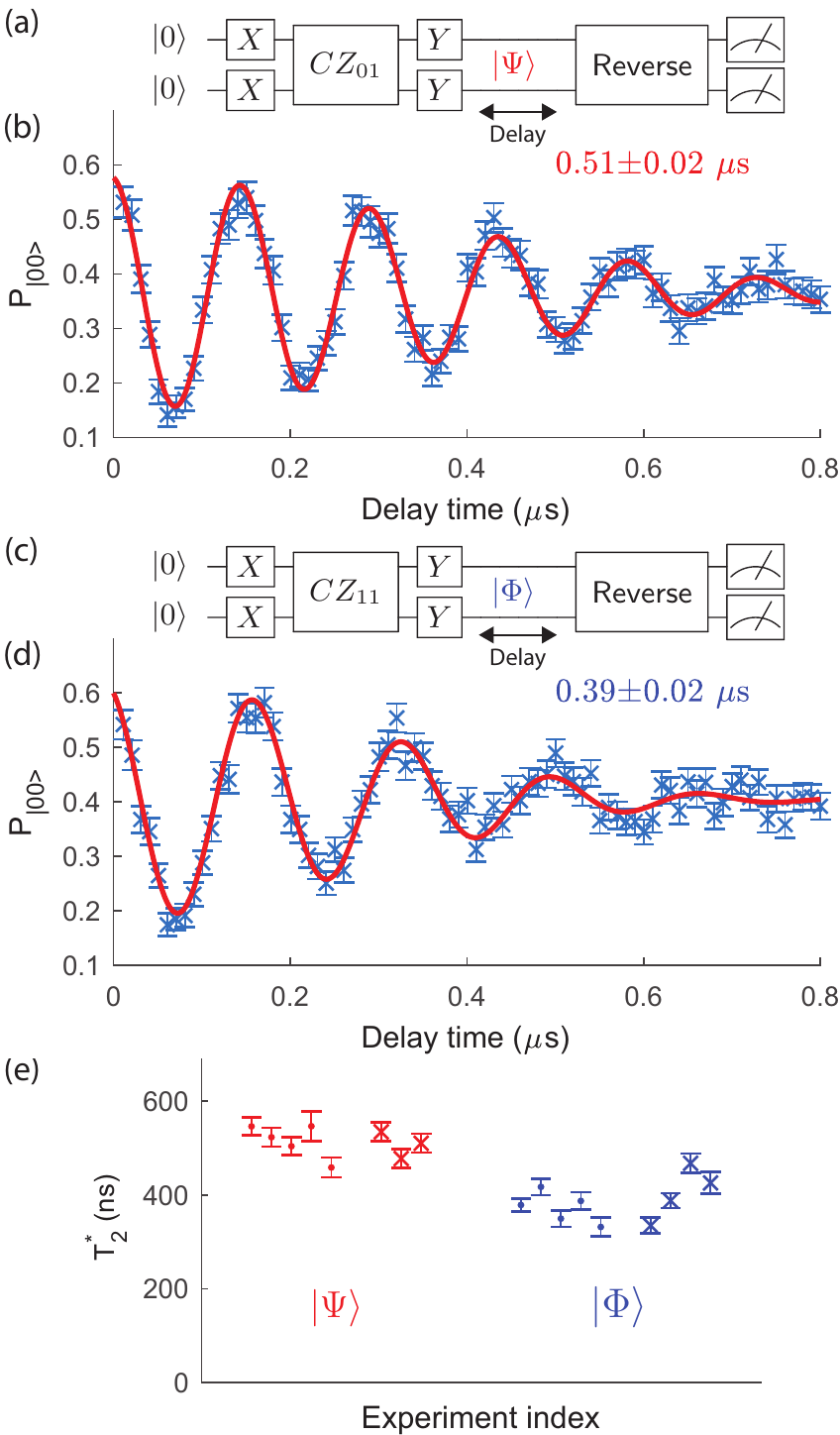}
	\caption{(a,c) Circuit diagrams for two-qubit experiments analogous to the measurement of Ramsey fringes. The gate sequences are designed such that single-qubit rotations are always applied simultaneously to both qubits, avoiding idle times that would lead to faster dephasing. Here $CZ_{ij}\ket{m,n} = (-1)^{\delta(i,m)\delta(j,n)}\ket{m,n}$ for $i,j,m,n \in \{0,1\}$~\cite{Watson2018}.
	(b,d) Typical $\ket{00}$ return probability as a function of delay time  for (b) $\ket{\Psi}$ and (d) $\ket{\Phi}$. The data are fit with a sinusoidal function with Gaussian decay, $P_{\ket{00}} \propto e^{-\left(t/T_2^*\right)^2}$. Error bars are based on a Monte Carlo method by assuming a multinomial distribution for the measured two-spin probabilities and are $\pm 1 \sigma$ from the mean~\cite{Watson2018}. (e) Scatter plot of decay times for $\ket{\Psi}$ and $\ket{\Phi}$ for two measurement runs separated by $\sim$50 hours (points and crosses). Every data point is averaged over $\sim$100 minutes. The average coherence times are $513 \pm 8$ ns and $387 \pm 6$ ns for $\ket{\Psi}$ and $\ket{\Phi}$, respectively. Error bars are $\pm 1 \sigma$ from the mean.}
	\label{fig:fig3}
\end{figure}

Concretely, we perform two-qubit measurements analogous to the measurement of Ramsey fringes to measure the decay of Bell state coherences over time~\cite{Premakumar2018}.
As shown in the circuits in Figs.~\ref{fig:fig3}(a,c), we prepare $\ket{\Psi}$ or $\ket{\Phi}$ and after a varying free evolution time we reverse the sequence to ideally return to the $\ket{00}$ state.
In every run of the experiment, we measure both spins in single-shot mode and determine the two-spin probabilities from repeated experiment runs. 
The two-spin probabilities are normalized and a Gaussian decay is fit to the $\ket{00}$ return probability.
To improve the fit of the decay, we add an evolution-time dependent phase to the first microwave pulse applied to Q2 after the delay time, so that the measured $\ket{00}$ probability oscillates. We first test the measurement procedure via artificially introduced dephasing from random rotations of each spin around its quantization axis, implemented in software via Pauli frame updates. As seen in Fig.~\ref{fig:Supp1} in the Supplemental Material~\cite{SuppMat}, the decay observed for the anti-parallel (parallel) Bell state is independent of the noise amplitude when the same (opposite) random rotations are applied to both spins, but increases when opposite (the same) random rotations are applied to the two spins, as expected. This validates the measurement protocol.

Figures~\ref{fig:fig3}(b,d) show typical decay curves for $\ket{\Psi}$ and $\ket{\Phi}$, respectively, when subject to natural noise only.
A scatter plot of repeated measurements, Fig.~\ref{fig:fig3}(e), shows a systematically longer $T_2^*$ for $\ket{\Psi}$ than for $\ket{\Phi}$, indicating correlations in the noise. 
Using Eq.~\ref{eq:T2s}, derived for quasi-static noise, we can extract from the decay of $\ket{\Psi}$ and $\ket{\Phi}$ a lower bound for the correlation factor, $\rho \geq 0.27\pm0.02$ (see Supplemental Material~\cite{SuppMat}).
In order to go beyond a lower bound and determine an estimate of $\rho$ from Eq.~\ref{eq:T2s}, we also need at least one of the single-qubit dephasing times, which we measured to be $T_{2,1}^* = 0.97\pm0.02 \,\mu$s and $T_{2,2}^* = 0.59\pm0.02 \,\mu$s. 
Using both single-qubit $T_2^*$s in Eq.~\ref{eq:T2s} gives an overdetermined system of equations. We proceed by keeping $T_{2,1}^*/T_{2,2}^*$ equal to the measured ratio, and obtain a modest correlation factor, $\rho = 0.31\pm0.03$ (see Supplemental Material~\cite{SuppMat}). In other experimental runs performed on the same sample, but separated in time by several months and with different gate voltage settings, we observed even smaller correlation factors.

We note that in keeping $T_{2,1}^*/T_{2,2}^*$ fixed, Eq.~\ref{eq:T2s} returns a value for $\sigma_1$ and $\sigma_2$ that is $\sim$15\% larger than the measured value. The discrepancy may be in part due to the fact that the simple model that leads to Eq.~\ref{eq:T2s} assumes quasi-static Gaussian noise. This is a commonly made assumption in simple models of silicon spin qubits, but various experiments showed higher frequency noise to be relevant as well \cite{Veldhorst2014,Yoneda2018,Watson2018}. 

\begin{figure}[!htb]
	\centering
	\includegraphics{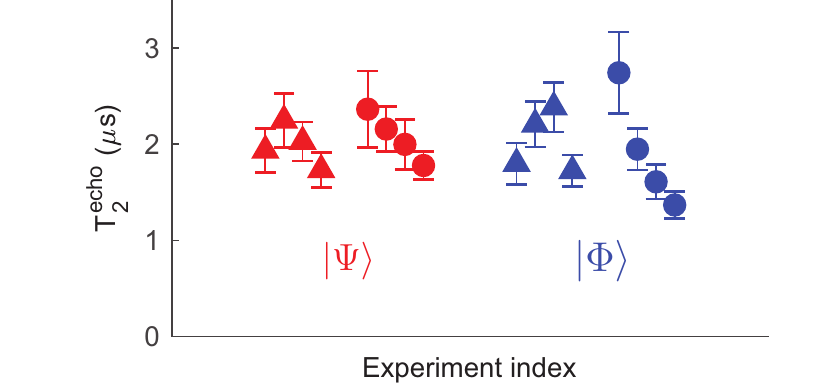}
	\caption{Scatter plot of the two-qubit coherence times obtained in Hahn-echo style measurements for $\ket{\Psi}$ and $\ket{\Phi}$, from a fit to the data with an exponentially decaying sinusoidal function ($P_{\ket{00}} \propto e^{-t/T_2^*}$). Triangles represent data points where the Hahn echo pulses applied to both qubits are rotations around the $\hat{x}$-axis. For the circles, the rotation of Q1 is around $\hat{x}$ and the rotation of Q2 is around $\hat{y}$. Data points are averaged over $\sim$[47, 66, 100, 148] minutes. The average two-qubit Hahn echo coherence times are $2.03 \pm 0.09 \, \mu$s and $1.98 \pm 0.09 \, \mu$s for $\ket{\Psi}$ and $\ket{\Phi}$, respectively. Error bars are $\pm1\sigma$ from the mean.}
	\label{fig:fig4}
\end{figure}

In order to gain insight into the frequency dependence of the spatial noise correlations, we perform measurements analogous to Hahn echo measurements. Here the delay times seen in the circuit diagrams of Fig.~\ref{fig:fig3}(a,c) contain 180 degree rotations around the $\hat{x}$ or $\hat{y}$ axis applied to the two qubits, which reverse the time evolution resulting from static noise contributions (see the Supplemental Material~\cite{SuppMat} for circuit diagrams and details).  
The results are presented in Fig.~\ref{fig:fig4}.
The echo pulses prolong the two-qubit coherence times by a factor of $\sim 4-5$.
We do not, however, observe a systematic difference in the echo decay times for the parallel versus anti-parallel Bell states, meaning there are no detectable spatial correlations in higher-frequency noise, and the correlations found in the Ramsey-style measurements of Fig.~\ref{fig:fig3} are mostly present in the low-frequency part of the spectrum.

To interpret the weak spatial correlations in the noise observed in the experiment, we first make a few observations.
Multiple independent noise sources that each produce perfectly correlated noise ($\rho=1$) acting with the same relative amplitude on the two qubits, are equivalent to a single (stronger) source of perfectly correlated noise acting with this same relative amplitude on the two qubits. However, the effect of multiple independent asymmetric, correlated noise sources acting with randomly distributed relative amplitudes on the two qubits, rapidly becomes indistinguishable from uncorrelated noise. This is illustrated in an example simulation of the combined effect of three asymmetric, correlated noise sources, shown in the Supplemental Material~\cite{SuppMat}.
As a more extreme example, the combination of perfectly correlated and perfectly anti-correlated noise with equal amplitude, is equivalent to  uncorrelated noise.
All of these effects are described by (see Supplemental Material~\cite{SuppMat}):
\begin{equation}
    \frac{T_{2,\ket{\Phi}}^*}{T_{2,\ket{\Psi}}^*} = \frac{\sigma_-}{\sigma_+} \propto \sqrt{\frac{\sum_i(\alpha_{i,1}-\alpha_{i,2})^2}{\sum_i(\alpha_{i,1}+\alpha_{i,2})^2}},
    \label{eq:T2ratio}
\end{equation}
where $\sigma_-$ and $\sigma_+$ are the standard deviations of the distributions of fluctuations in the difference and sum of the frequencies of the two qubits, respectively, and $\alpha_{i,j}$ is the coupling strength of noise source $i$ to qubit $j$.

We now discuss the effect of the known noise mechanisms acting on spin qubits and the expected spatial correlations for each mechanism. 
Fluctuating background charges in the substrate, interfaces or dielectrics directly affect the qubit splitting because of the magnetic field gradient produced by the micromagnets. 
When these charges are located close to the dots, they will generally couple differently to the two qubits, introducing asymmetric noise. Specifically for charge fluctuators located in between the two dots, even anti-correlated noise may result. For distant charges, the coupling becomes more symmetric, but several factors can lead to asymmetric noise amplitudes even in this case, for instance a difference in the confining potential between the two dots or a difference in the strength of the local magnetic field gradient. We have clear evidence of a pronounced difference in the confining potential of the two dots in this sample, based on the sensitivity of the respective qubit splittings to changes in gate voltages (see Supplemental Material~\cite{SuppMat}). 
Similar considerations apply to the effect of gate voltage noise, which also couples to the qubit splitting through the magnetic field gradient.
Another important noise source in this natural silicon substrate is hyperfine interaction with nuclear spins, for which little or no spatial correlations are expected~\cite{Chekhovich2013}. 

Our expectations for the spatial noise correlations based on our understanding of the system physics are consistent with the experimental results and our theoretical observations on the combined effect of multiple noise sources. A picture emerges where noise from  multiple distant charge fluctuators that affect the qubits asymmetrically due to their different confining potentials, is responsible for the (weak) spatial noise correlations at low frequency. Additional uncorrelated noise is introduced by the coupling to the nuclear spins.

In summary, we have presented an experimental study of spatial noise correlations based on the coherence of Bell states in a Si/SiGe two-qubit device.
Experimentally we observe small spatial correlations in low-frequency noise, while for higher-frequency noise correlations appear to be absent.
Our findings on the importance of asymmetric coupling of noise sources to two (or more) qubits can be exploited for reducing or enhancing spatial correlations in the noise in any qubit platform. For the case of spin qubits in quantum dots, this can be done for instance through a device design with engineered differences in confining potential or magnetic field gradient. In this respect, qubits encoded in two-electron spin states in dot-donor systems offer an extreme difference in confining potential~\cite{Harvey-Collard2017b}. 
We anticipate that the optimization of future quantum error correction codes will go hand in hand with the design of qubits that either maximize or minimize spatial noise correlations.

Data supporting the findings of this study are available online \cite{Zenodo_Noise}.


\begin{acknowledgments}
The authors acknowledge useful discussions with the members of the Vandersypen group, software support by F. van Riggelen, and technical assistance by M.~L.~I. Ammerlaan, O.~W.~B. Benningshof, J.~H.~W. Haanstra, J.~D. Mensingh, R.~G. Roeleveld, R.~A. Schoonenboom, R.~N. Schouten, M.~J. Tiggelman, R.~F.~L. Vermeulen and S. Visser.
We acknowledge financial support by Intel Corporation.
Development and maintenance of the growth facilities used for fabricating samples is supported by DOE (DE-FG02-03ER46028).
We acknowledge the use of facilities supported by NSF through the University of Wisconsin-Madison MRSEC (DMR-1121288).
Research was sponsored by the Army Research Office (ARO), and was accomplished under Grant Number W911NF-17-1-0274. The views and conclusions contained in this document are those of the authors and should not be interpreted as representing the official policies, either expressed or implied, of the Army Research Office (ARO), or the U.S. Government. The U.S. Government is authorized to reproduce and distribute reprints for Government purposes notwithstanding any copyright notation herein.
\end{acknowledgments}

%

\clearpage
\newpage

	
\renewcommand{\thefigure}{S\arabic{figure}}
\renewcommand{\thetable}{S\arabic{table}}
\renewcommand{\theequation}{S\arabic{equation}}
\renewcommand{\thesection}{\Roman{section}} 
\renewcommand{\thesubsection}{\Alph{subsection}}
\setcounter{figure}{0}
\setcounter{equation}{0}
\pagenumbering{roman}

\maketitle

\onecolumngrid
\begin{centering}
{\Large Supplemental Material for} \\
\vspace{2mm}

{\large \textbf{Spatial Noise Correlations in a Si/SiGe Two-Qubit Device from Bell State Coherences}}\\
\vspace{4mm}

{\normalsize Jelmer M. Boter,$^{1,*}$ X. Xue,$^{1,*}$ Tobias S. Kr\"ahenmann,$^1$ Thomas F. Watson,$^1$ \\
Vickram N. Premakumar,$^2$ Daniel R. Ward,$^2$ Donald E. Savage,$^2$ Max G. Lagally,$^2$ Mark Friesen,$^2$ \\
Susan N. Coppersmith,$^{2,\dagger}$ Mark A. Eriksson,$^2$ Robert Joynt,$^2$ and Lieven M.~K. Vandersypen$^{1,3,\ddagger}$}\\
\vspace{2mm}

\small{$^{1}$\textit{QuTech and Kavli Institute of Nanoscience, Delft University of Technology, Lorentzweg 1, 2628 CJ Delft, The Netherlands}}\\
\small{$^{2}$\textit{University of Wisconsin-Madison, Madison, WI 53706, USA}}\\
\small{$^{3}$\textit{Components Research, Intel Corporation, 2501 NE Century Blvd, Hillsboro, OR 97124, USA}}\\
\vspace{1mm}

\small{$^{*}$These authors contributed equally to this work.}\\
\small{$^{\dagger}$Present address: School of Physics, University of New South Wales, Sydney NSW 2052, Australia.}\\
\small{$^{\ddagger}$To whom correspondence should be addressed: \href{mailto://l.m.k.vandersypen@tudelft.nl}{l.m.k.vandersypen@tudelft.nl}}

\end{centering}

\section{Measurement setup}
The measurement setup used in this work is the same as the setup used by \citet{Watson2018} and \citet{Xue2019}.
The measurements were done at a temperature of $T\approx30$ mK in an external magnetic field of B\textsubscript{ext} = 617 mT.
DC voltages are set via filtered lines from room-temperature digital-to-analog converters.
Tektronix 5014C arbitrary waveform generators (AWGs) are connected to gates P1 and P2 via coaxial cables for gate voltage pulses.
Keysight E8267D vector microwave sources are connected to gates MW1 and MW2 for EDSR.
I/Q input channels of the microwave sources are connected to a master AWG to control frequency, phase and duration of the microwave bursts via I/Q modulation.
The phase of the microwave drive signal determines the rotation axis in the $\hat{x}-\hat{y}$ plane of the Bloch sphere, and we update the rotating reference frame in software to perform $\hat{z}$ rotations \cite{Vandersypen2005}.
Pulse modulation is used to increase the on/off ratio of the microwave bursts.
The master AWG also controls the clock of the entire system and triggers all the other instruments.
Data acquisition is done by a Spectrum M4i.44 digitizer card that is installed in the measurement computer.
This card records the sensing dot current traces at a sampling rate of $\sim$60 kHz after passing through a 12-kHz Bessel low-pass filter (SIM965).
Threshold detection is used to convert each trace to a single bit value (0 or 1) by the measurement computer.

\section{Qubit characteristics}
\begin{table}[!hbt]
	\centering
    \begin{tabular}{c|c|c}
        & Q1 & Q2 \\
        \hline
        f & 18.35 GHz & 19.61 GHz \\
        $T_1$ & $>$50 ms \cite{Watson2018} & 3.7$\pm$0.5 ms \cite{Watson2018} \\
        $T^{*}_2$ & 0.97$\pm$0.02 $\mu$s & 0.59$\pm$0.02 $\mu$s \\
        $T_2^{Hahn}$ & 6.8$\pm$0.3 $\mu$s & 2.8$\pm$0.2 $\mu$s \\
        $F_{\ket{\Psi^+}}$ & \multicolumn{2}{c}{0.88$\pm$0.02 \cite{Watson2018}} \\
        $F_{\ket{\Psi^-}}$ & \multicolumn{2}{c}{0.88$\pm$0.02 \cite{Watson2018}} \\
        $F_{\ket{\Phi^+}}$ & \multicolumn{2}{c}{0.85$\pm$0.02 \cite{Watson2018}} \\
        $F_{\ket{\Phi^-}}$ & \multicolumn{2}{c}{0.89$\pm$0.02 \cite{Watson2018}} \\
    \end{tabular}
    \caption{Relevant single-qubit characteristics for simultaneous driving of both qubits, and Bell state fidelities F for the four Bell states. All errors are $\pm1\sigma$ from the mean.}
    \label{tab:qubit_char}
\end{table}
\noindent An upper bound on the residual exchange during single-qubit gates and free evolution of 100 kHz is determined, using the methods of \citet{Watson2018}.

\section{Noise model}
\noindent We model the two-qubit system by the Hamiltonian:
\begin{equation}
    H =  \frac{hf_1}{2}\sigma_1^Z + \frac{hf_2}{2}\sigma_2^Z,
\end{equation}
where $h$ is the Planck constant, $f_i = \frac{g\mu_B B_i}{h}$ is the Larmor frequency for qubit $i$, $g$ is the electron g-factor, $\mu_B$ is the Bohr magneton, $B_i$ is the total magnetic field at the position of qubit $i$ and $\sigma_i^Z$ is the Pauli Z operator for qubit $i$.
The two qubits are subject to dephasing noise, which we model as a fluctuating qubit frequency $f_i$.
We assume Gaussian distributed noise with zero mean and covariance matrix $\Sigma$:
\begin{equation}
    \textbf{f} = (f_1,f_2) \sim \mathcal{N}((0,0),\Sigma); \,
    \Sigma = 
    \begin{bmatrix}
    	\sigma_1^2 & \rho\sigma_1\sigma_2 \\
        \rho\sigma_1\sigma_2 & \sigma_2^2
	\end{bmatrix},
\end{equation}
where $\sigma_i^2$ is the variance of the noise in $f_i$, and $\rho$ ($-1 \leq \rho \leq 1$) is a correlation factor.
Positive $\rho$ indicates correlations, while negative $\rho$ indicates anti-correlations.
We obtain the unitary time evolution operator by exponentiating the Hamiltonian:
\begin{equation}
    U = e^{-iHt/\hbar} = 
    \begin{pmatrix}
        e^{-i\pi(f_1+f_2)t} & & & \\
        & e^{-i\pi(f_1-f_2)t} & & \\
        & & e^{i\pi(f_1-f_2)t} & \\
        & & & e^{i\pi(f_1+f_2)t}
    \end{pmatrix},
\end{equation}
where $\hbar = \frac{h}{2\pi}$.
Assuming quasi-static noise, we average over this unitary transformation by integrating over the joint probability distribution function:
\begin{equation}
    \rho(t) = \overline{U\rho(0)U^\dagger} = \frac{1}{2\pi\sqrt{\textrm{det}(\Sigma)}} \int U\rho(0)U^\dagger e^{-\textbf{f}^T\Sigma^{-1}\textbf{f}/2} d\textbf{f}.
\end{equation}
The relevant expressions for anti-parallel ($\ket{\Psi}$) and parallel ($\ket{\Phi}$) Bell states are:
\begin{equation}
	\begin{split}
    \bra{01}\overline{U\rho(0)U^\dagger}\ket{10} &= \frac{1}{2} \times \frac{1}{2\pi\sqrt{\textrm{det}(\Sigma)}} \int e^{-i2\pi(f_1-f_2)t} e^{-\textbf{f}^T\Sigma^{-1}\textbf{f}/2} d\textbf{f} = \frac{1}{2}\textrm{exp}\left[-2\pi^2t^2(\sigma_1^2 + \sigma_2^2 - 2\rho\sigma_1\sigma_2) \right], \\
    \bra{00}\overline{U\rho(0)U^\dagger}\ket{11} &= \frac{1}{2} \times \frac{1}{2\pi\sqrt{\textrm{det}(\Sigma)}} \int e^{-i2\pi(f_1+f_2)t} e^{-\textbf{f}^T\Sigma^{-1}\textbf{f}/2} d\textbf{f} = \frac{1}{2}\textrm{exp}\left[-2\pi^2t^2(\sigma_1^2 + \sigma_2^2 + 2\rho\sigma_1\sigma_2) \right],
    \end{split}
\end{equation}
so the decay for anti-parallel and parallel Bell states is Gaussian with associated time scales (Eq.~\ref{eq:T2s} of the main text):
\begin{equation}
	\begin{split}
    \left(\frac{1}{T_{2,\ket{\Psi}}^*}\right)^2 & = 2\pi^2 \left(\sigma_1^2 + \sigma_2^2 - 2\rho\sigma_1\sigma_2 \right), \\
    \left(\frac{1}{T_{2,\ket{\Phi}}^*}\right)^2 & = 2\pi^2 \left(\sigma_1^2 + \sigma_2^2 + 2\rho\sigma_1\sigma_2 \right).
    \end{split}
    \label{eq:T2s_Supp}
\end{equation}
Noting that in the case of Gaussian quasi-static noise for single-qubit decay $\left(\frac{1}{T_{2,i}^*}\right)^2 = 2\pi^2\sigma_i^2$, these expressions can be rewritten in terms of single-qubit coherence times:
\begin{align}
    \left(\frac{1}{T_{2,\ket{\Psi}}^*}\right)^2 & = \left(\frac{1}{T_{2,1}^*}\right)^2 + \left(\frac{1}{T_{2,2}^*}\right)^2 - 2\rho\frac{1}{T_{2,1}^*T_{2,2}^*},
    \label{eq:T2Psi}\\
    \left(\frac{1}{T_{2,\ket{\Phi}}^*}\right)^2 & = \left(\frac{1}{T_{2,1}^*}\right)^2 + \left(\frac{1}{T_{2,2}^*}\right)^2 + 2\rho\frac{1}{T_{2,1}^*T_{2,2}^*}.
    \label{eq:T2Phi}
\end{align}
Subtracting Eq.~\ref{eq:T2Psi} from Eq.~\ref{eq:T2Phi}, we express the correlation factor $\rho$ in terms of the single- and two-qubit coherence times as:
\begin{equation}
	\rho = \frac{T_{2,1}^*T_{2,2}^*}{4} \left[\left(\frac{1}{T_{2,\ket{\Phi}}^*}\right)^2 - \left(\frac{1}{T_{2,\ket{\Psi}}^*}\right)^2 \right].
    \label{eq:rho3}
\end{equation}
In addition, Eqs.~\ref{eq:T2Psi} and \ref{eq:T2Phi} allow one to formulate a sum rule, and to define a violation parameter $\Delta_s$ (with dimensions of a rate) that quantifies the difference between the model and experimental results:
\begin{equation}
	\left(\frac{1}{T_{2,\ket{\Psi}}^*}\right)^2 + \left(\frac{1}{T_{2,\ket{\Phi}}^*}\right)^2 - 2\left[ \left(\frac{1}{T_{2,1}^*}\right)^2 + \left(\frac{1}{T_{2,2}^*}\right)^2 \right] = \Delta_s^2,
	\label{eq:Delta_s}
\end{equation}
In this work, the four coherence times $T_{2,1}^*$, $T_{2,2}^*$, $T_{2,\ket{\Psi}}$ and $T_{2,\ket{\Phi}}^*$ are obtained individually and for matching model and experimental data $\Delta_s = 0$.

The model presented before assumes quasi-static noise.
Without presenting the details here, a similar sum rule with corresponding violation parameter $\Delta_f$ can be obtained for non-quasi-static noise:
\begin{equation}
	\frac{1}{T_{2,\ket{\Psi}}^*} + \frac{1}{T_{2,\ket{\Phi}}^*} - 2\left(\frac{1}{T_{2,1}^*} + \frac{1}{T_{2,2}^*}\right) = \Delta_f.
	\label{eq:Delta_f}
\end{equation}
The violation parameter $\Delta_f$ is based on a sum rule for non-quasi-static noise, so coherence times obtained from exponential fits ($P_{\ket{00}} \propto e^{-t/T_2^*}$) have to be used in this expression.

\section{Method verification}
\begin{figure}[!htb]
	\centering
	\includegraphics[width=0.5\textwidth]{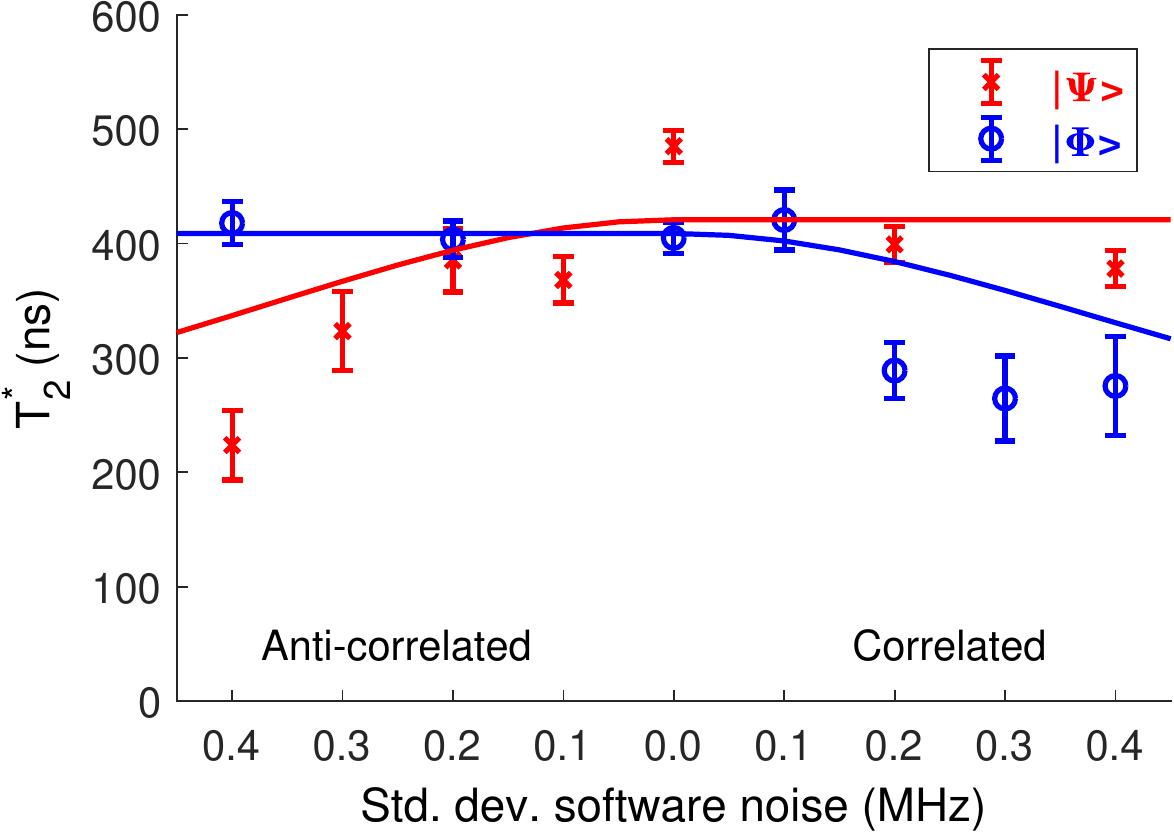}
	\caption{Bell state coherence times for added software noise. The horizontal axis represents the amplitude of added noise in units of frequency.
In the origin no noise is added, so this measurement only includes naturally existing noise.
Moving to the right (left) on the horizontal axis, we increasingly add (anti-)correlated noise.
Data points result from a fit to Gaussian decay, $P_{\ket{00}} \propto e^{-\left(t/T_2^*\right)^2}$. Error bars are $\pm1\sigma$ from the mean.
The solid lines represent a theoretical prediction.}
	\label{fig:Supp1}
\end{figure}

To verify the method used in this work, we inject artificial noise in the experiments by applying random software Z rotations to the qubits and measure the coherence times for $\ket{\Phi}$ and $\ket{\Psi}$.
These rotations are implemented by adding an evolution time dependent phase to the first microwave pulse after the waiting time, in addition to the phase to improve the fit of the decay.
The frequency fluctuations corresponding to this extra phase are sampled from a Gaussian distribution with varying standard deviation.
Adding (anti-)correlated noise is expected to have an effect on $\ket{\Phi}$ ($\ket{\Psi}$), but not on $\ket{\Psi}$ ($\ket{\Phi}$). The results of this control experiment are shown in Fig.~\ref{fig:Supp1}. Despite the large error bars on some of the data points, we clearly observe the expected trend, showing the method to be reliable

\section{Bell state fidelity}
The Bell state fidelity does affect the method used in this work.
Imperfect Bell state initialization results in a finite amplitude for other off-diagonal elements of the density matrix than those of interest, which mixes in decays with different characteristic time scales.
In the present experiments, the Bell states have not been characterized, but from the Bell state density matrices presented in the Supplementary Information of \citet{Watson2018}, we estimate the amplitudes of the not-intended density matrix elements to not exceed $\sim$19\% of the elements of interest and most of them are much smaller.
In our experiments we do not see clear deviation from a single decay.

\section{Quantifying correlations}
In case the experimental data is not fully consistent with the simple quasi-static model, as quantified by the violation parameter $\Delta_s$ in Eq.~\ref{eq:Delta_s}, it is still possible to use this model to extract quantitative information on the correlations in the noise acting on the qubits based only on the two-qubit coherence times.
From Eqs.~\ref{eq:T2Psi} and \ref{eq:T2Phi}, given the two-qubit coherence times, effective single-qubit coherence times can be calculated as:
\begin{equation}
    \left(\frac{1}{T_{2,1(2)}^*}\right)^2 = \frac{\left(\frac{1}{T_{2,\ket{\Phi}}^*}\right)^2 + \left(\frac{1}{T_{2,\ket{\Psi}}^*}\right)^2}{4} \mp \frac{1}{2}\sqrt{\left( \frac{\left(\frac{1}{T_{2,\ket{\Phi}}^*}\right)^2 + \left(\frac{1}{T_{2,\ket{\Psi}}^*}\right)^2}{2} \right)^2 - 4\left( \frac{\left(\frac{1}{T_{2,\ket{\Phi}}^*}\right)^2 - \left(\frac{1}{T_{2,\ket{\Psi}}^*}\right)^2}{4\rho} \right)^2},
    \label{eq:T2_eff}
\end{equation}
where the minus (plus) sign corresponds to Q1 (Q2), assuming $T_{2,1}^* \geq T_{2,2}^*$.
Solutions only exist if the argument of the square root is equal to or larger than zero, so for
\begin{equation}
    |\rho| \geq \rho_{min} = \left| \frac{\left(\frac{1}{T_{2,\ket{\Phi}}^*}\right)^2 - \left(\frac{1}{T_{2,\ket{\Psi}}^*}\right)^2}{\left(\frac{1}{T_{2,\ket{\Phi}}^*}\right)^2 + \left(\frac{1}{T_{2,\ket{\Psi}}^*}\right)^2} \right|.
\end{equation}
Using this simple model, we find a lower bound for the correlation factor $\rho_{min} = 0.27\pm0.02$.

Taking into account the experimental single-qubit coherence times and assuming their ratio ($\beta = \frac{T^*_{2,2}}{T^*_{2,1}}$) to be fixed, effective single-qubit coherence times can be obtained by adding Eqs.~\ref{eq:T2Psi} and \ref{eq:T2Phi}, and are given by:
\begin{equation}
    \left(\frac{1}{T_{2,1}^*}\right)^2 = \left(\frac{\beta}{T_{2,2}^*}\right)^2 = \frac{\beta^2}{2(1+\beta^2)} \left[ \left(\frac{1}{T_{2,\ket{\Phi}}^*}\right)^2 + \left(\frac{1}{T_{2,\ket{\Psi}}^*}\right)^2 \right].
\end{equation}
The correlation factor $\rho$ from Eq.~\ref{eq:rho3} in that case is expressed as:
\begin{equation}
	\rho = \frac{\beta \left(T_{2,1}^*\right)^2}{4} \left[ \left(\frac{1}{T_{2,\ket{\Phi}}^*}\right)^2 - \left(\frac{1}{T_{2,\ket{\Psi}}^*}\right)^2 \right].
	\label{eq:rho_3_fixed_ratio}
\end{equation}
For the experimental value $\beta = 0.61\pm0.02$ ($T_{2,1}^* = 0.97\pm0.02 \,\mu$s and $T_{2,2}^* = 0.59\pm0.02 \,\mu$s), we find a correlation factor $\rho = 0.31\pm0.03$, and effective single-qubit coherence times $T_{2,1}^* = 0.84\pm0.03 \, \mu$s and $T_{2,2}^* = 0.51\pm 0.02 \, \mu$s.

\section{Echo experiments}
Dynamical decoupling sequences can be used to investigate the frequency dependence of spatial noise correlations, similar to mapping out the frequency spectrum of noise acting on a single qubit~\cite{Muhonen2014,Kawakami2016,Yoneda2018}.
\begin{figure}[!htb]
	\centering
	\includegraphics{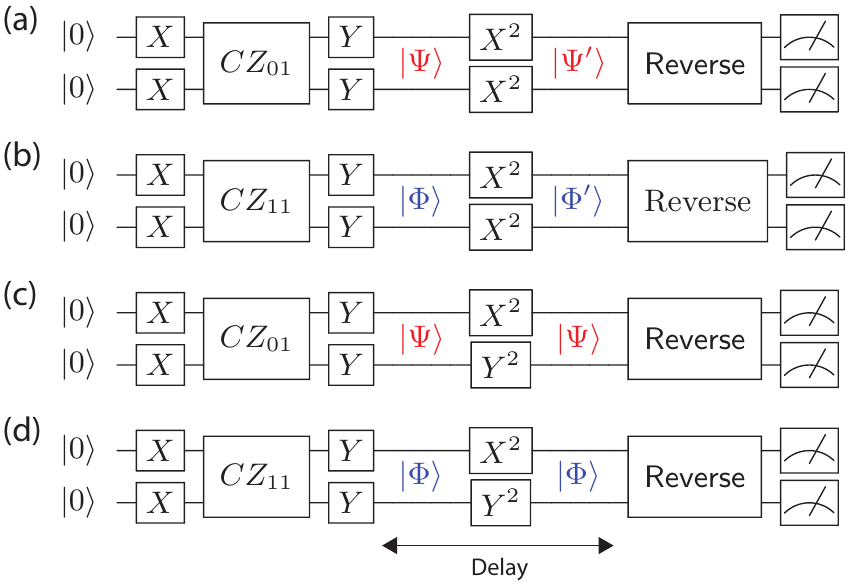}
	\caption{Circuit diagrams for two different versions (XX (a,b) and XY (c,d)) of an experiment analogous to the measurement of a Hahn echo for $\ket{\Psi}$ (a,c) and $\ket{\Phi}$ (b,d).}
	\label{fig:Supp2}
\end{figure}
In addition to the measurements analogous to Ramsey experiments, we performed measurements analogous to a Hahn echo experiment with a single decoupling pulse on each qubit halfway the waiting time.
Results are presented in Fig.~\ref{fig:fig4} of the main text.
We performed two versions of the echo experiment to which we refer as XX and XY echo, respectively.
In the XX echo experiment we apply a $\pi_X$ pulse on both qubits, which transforms $\ket{\Psi}  = (\ket{\downarrow\uparrow} -i \ket{\uparrow\downarrow})/\sqrt{2}$ into $\ket{\Psi'}  = (\ket{\downarrow\uparrow} +i \ket{\uparrow\downarrow})/\sqrt{2}$, and $\ket{\Phi} = (\ket{\downarrow\downarrow} -i \ket{\uparrow\uparrow})/\sqrt{2}$ into $\ket{\Phi'} = (\ket{\downarrow\downarrow} +i \ket{\uparrow\uparrow})/\sqrt{2}$, as shown in the circuits in Figs.~\ref{fig:Supp2}(a,b).
The XY echo experiment consists of a $\pi_X$ pulse on Q1 and a $\pi_Y$ pulse on Q2, which transforms $\ket{\Psi}$ and $\ket{\Phi}$ to itself, as shown in the circuits in Figs.~\ref{fig:Supp2}(c,d). The difference between the XX and XY sequences is analogous to that between single-qubit echo pulses around $\hat{x}$ versus $\hat{y}$. 
We do note that for both versions of the two-qubit decoupling used in this work, the two-qubit state is taken out of the logical qubit space during the pulses.

\section{Adding multiple independent noise sources}
To derive Eq.~\ref{eq:T2ratio} of the main text, consider a single noise source $i$ with coupling strength $\alpha_{i,1}$ and $\alpha_{i,2}$ (which can be expressed for instance in units of MHz/mV, if noise source $i$ is expressed in units of mV) to qubit 1 and qubit 2 respectively.
The noise source fluctuates with standard deviation $\sigma$.
The standard deviations of the fluctuations in the difference ($f_1-f_2$) and sum ($f_1+f_2$) of the frequencies are then given by:
\begin{equation}
	\begin{split}
		\sigma_{i,-} &= \sigma|\alpha_{i,1}-\alpha_{i,2}|, \\
		\sigma_{i,+} &= \sigma|\alpha_{i,1}+\alpha_{i,2}|. \\
	\end{split}
	\label{eq:sigma_dif_sum}
\end{equation}
For $N$ independent noise sources the combined standard deviation is given by:
\begin{equation}
	\sigma^2 = \Sigma_i^N \sigma_i^2.
	\label{eq:sigma_sum}
\end{equation}
Combining Eqs.~\ref{eq:sigma_dif_sum} and \ref{eq:sigma_sum} gives:
\begin{equation}
	\begin{split}
		\sigma_- = \sqrt{\Sigma_i^N \sigma_{i,-}^2} = \sigma\sqrt{\Sigma_i^N \left(\alpha_{i,1}-\alpha_{i,2}\right)^2}, \\
		\sigma_+ = \sqrt{\Sigma_i^N \sigma_{i,-}^2} = \sigma\sqrt{\Sigma_i^N \left(\alpha_{i,1}+\alpha_{i,2}\right)^2},
	\end{split}
\end{equation}
where we absorb differences in standard deviations between noise sources in the coupling strengths.
Since $T_2^* \propto \frac{1}{\sigma}$, this yields Eq.~\ref{eq:T2ratio} of the main text:
\begin{align}
    \frac{T_{2,\ket{\Phi}}^*}{T_{2,\ket{\Psi}}^*} &= \frac{\sigma_-}{\sigma_+} \propto \sqrt{\frac{\sum_i(\alpha_{i,1}-\alpha_{i,2})^2}{\sum_i(\alpha_{i,1}+\alpha_{i,2})^2}}.
\end{align}

\newpage
\section{Simulation of multiple asymmetric noise sources}
\begin{figure}[!htb]
	\centering
	\includegraphics{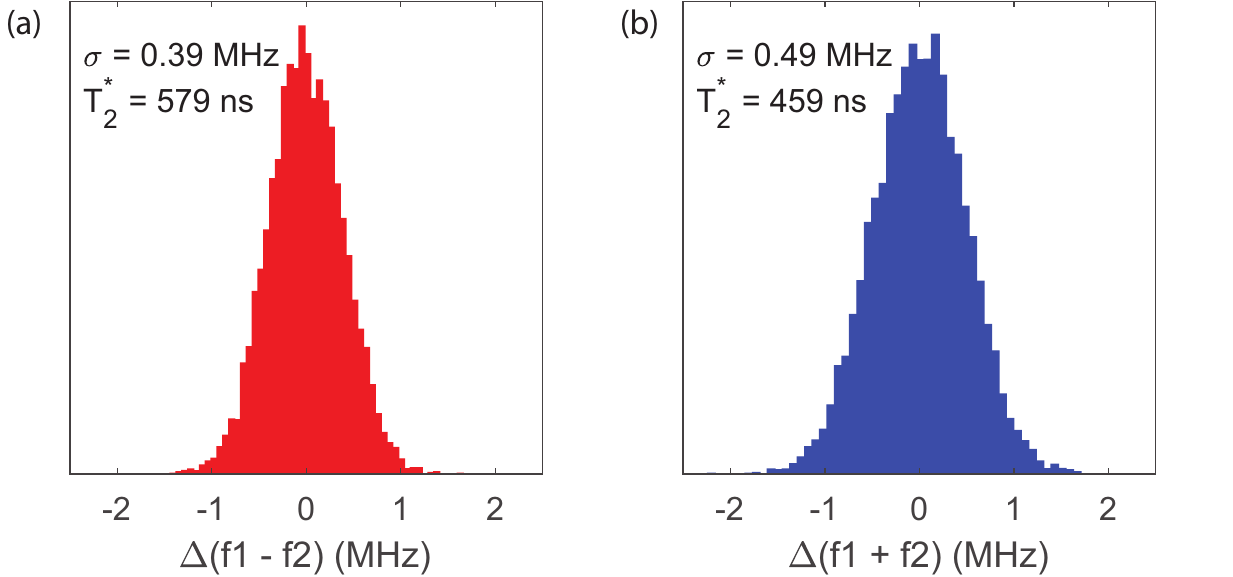}
	\caption{Simulation of three noise sources with coupling factors chosen to correspond to the experimentally measured coupling factors for three of the gate electrodes on the sample, namely P1, P2 and MW2 in Fig.~\ref{fig:fig1}(a) of the main text. The coupling factors to the two qubits for these and five other gate electrodes are tabulated in Table~\ref{tab:coupling_factors}. For all three gate electrodes, voltage fluctuations are sampled from a Gaussian distribution with 50 $\mu$V standard deviation. After sampling gate voltage fluctuations, the corresponding total frequency fluctuations for both qubits, and their difference and sum are calculated. The distributions of the fluctuations in (a) difference and (b) sum frequency are plotted.}
	\label{fig:Supp3}
\end{figure}

The result of a simulation of the combined effect of three asymmetric, correlated noise sources is shown in Fig.~\ref{fig:Supp3}.
The standard deviations of the distributions of fluctuations in difference and sum frequencies indicate that only modest correlations in the noise remain for their combined effect.

\begin{table}[!hbt]
	\centering
    \begin{tabular}{c|c|c}
        & Q1 & Q2 \\
        \hline
        P1 & -1 & -2 \\
        P2 & 0.175 & 0.8 \\
        MW1 & -0.015 & 0.025 \\
        MW2 & 0.8 & 8.5 \\ 
        B & 0.43 & 0.36 \\
        LD & -0.1 & -1.44 \\
        accQD & 0.9 & -1.8 \\
        accRes & -0.8 & -3.75 \\
    \end{tabular}
    \caption{Coupling factors (in MHz/mV) of eight of the surface gate electrodes on our sample to the two qubits.}
    \label{tab:coupling_factors}
\end{table}

\end{document}